\documentclass[12pt] {article}
\usepackage{psfrag}
\usepackage{graphicx}
\usepackage{latexsym,amsfonts}
\usepackage{amssymb}
\begin{document}
\setcounter{page}{1}

\title{Quantum field theoretic model of metastable resonant
spin--singlet state of the $np$ pair}

\author{A. N. Ivanov\,\thanks{E--mail: ivanov@kph.tuwien.ac.at, Tel.:
+43--1--58801--14261, Fax: +43--1--58801--14299}~\thanks{Permanent
Address: State Polytechnic University, Department of Nuclear Physics,
195251 St. Petersburg, Russian Federation}\,,
M. Cargnelli\,\thanks{E--mail: michael.cargnelli@oeaw.ac.at}\,,
M. Faber\,\thanks{E--mail: faber@kph.tuwien.ac.at, Tel.:
+43--1--58801--14261, Fax: +43--1--58801--14299}\,, H.
Fuhrmann\,\thanks{E--mail: hermann.fuhrmann@oeaw.ac.at}\,,\\
V. A. Ivanova\,\thanks{E--mail: viola@kph.tuwien.ac.at, State
Polytechnic University, Department of Nuclear Physics, 195251
St. Petersburg, Russian Federation}\,, J. Marton\,\thanks{E--mail:
johann.marton@oeaw.ac.at}\,, N. I. Troitskaya\,\thanks{State
Polytechnic University, Department of Nuclear Physics, 195251
St. Petersburg, Russian Federation}~~, J. Zmeskal\,\thanks{E--mail:
johann.zmeskal@oeaw.ac.at}}

\date{\today}

\maketitle

\begin{center}
{\it Atominstitut der \"Osterreichischen Universit\"aten,
Arbeitsbereich Kernphysik und Nukleare Astrophysik, Technische
Universit\"at Wien, \\ Wiedner Hauptstr. 8-10, A-1040 Wien,
\"Osterreich \\ und\\ Institut f\"ur Mittelenergiephysik
\"Osterreichische Akademie der Wissenschaften,\\
Boltzmanngasse 3, A-1090, Wien, \"Osterreich}
\end{center}

\begin{center}
\begin{abstract}
The $np$ pair in the spin--singlet state is treated as a Cooper
$np$--pair within the extension of the Nambu--Jona--Lasinio model of
light nuclei, describing the deuteron as a Cooper $np$--pair in the
spin--triplet state. For the Cooper $np$--pair in the spin--singlet
state we compute the binding energy and express the S--wave scattering
length of $np$ scattering in the spin--singlet state in terms of the
binding energy. The theoretical value of the S--wave scattering length
of $np$ scattering agrees well with the experimental data.
\end{abstract}


\end{center}

\newpage

At very low energies the S--wave amplitude of $np$ scattering in the
${^1}{\rm S}_0$ state (the spin--singlet state) can be written as
\cite{LL65,TE88}
\begin{eqnarray}\label{label1}
f_{np}(k) = \frac{1}{\displaystyle - \kappa_{np} - i\,k},
\end{eqnarray}
where $1/\kappa_{np} = a_{np}$ can be identified with the S--wave
scattering length of $np$ scattering in the spin--singlet
state\,\footnote{For simplicity we neglect the contribution of the
effective range \cite{LL65,TE88}.} and $k$ is a relative momentum of
the $np$ pair. The experimental value of the S--wave scattering
lengths of $np$ scattering in the spin--singlet state is equal to
$a_{np} = (-\,23.75\pm 0.01)\,{\rm fm}$ \cite{TE88}.

According to Landau and Lifshitz \cite{LL65}, the S--wave scattering
length $a_{np} = (-\,23.75\pm 0.01)\,{\rm fm}$ of $np$ scattering in
the spin--singlet state testifies the existence of the {\it virtual
level} with the binding energy $\varepsilon_{{^1}{\rm S}_0} =
\kappa^2_{np}/m_N = 1/a^2_{np}m_N = 0.074\,{\rm MeV}$,
where $m_N = 940\,{\rm MeV}$ is the nucleon mass.

It is well--known that strong low--energy interactions of the $np$
pair in the ${^3}{\rm S}_1$ state (the spin--triplet state) lead to
the existence of the bound state, the deuteron \cite{TE88} with the
binding energy $\varepsilon_D = 2.225\,{\rm MeV}$ \cite{TE88}. The
quantum field theoretic approach to the description of the deuteron as
the Cooper $np$ pair has been developed within the
Nambu--Jona--Lasinio model of light nuclei in Refs.
\cite{AI1}--\cite{AI5}\,\footnote{We use the abbreviation the NNJL
model that means the Nuclear Nambu--Jona--Lasinio model \cite{AI1}.}.

As has been shown in \cite{AI1}--\cite{AI5} the NNJL model describes
well the low--energy parameters of the deuteron, such as the binding
energy, the dipole magnetic and electric quadrupole moments
\cite{AI1}, the $\Delta\Delta$ component of the wave function of the
deuteron in the ground state \cite{AI2}, the low--energy nuclear
reactions with the deuteron of astrophysical interest \cite{AI3}, the
asymptotic ratio $D/S$ \cite{AI4} of the wave function of the
deuteron\,\footnote{The asymptotic ratio $D/S$ of the D--wave
component to the S--wave component of the wave function of the
deuteron in the ground state has been computed in the NNJL model in
agreement with the results obtained by Ericson within the potential
model approach \cite{TE82} and the experimental value, which has been
used by Kamionkowski and Bahcall \cite{TE82} for the calculation of
the astrophysical factor for the solar proton burning $p + p \to d +
e^+ + \nu_e$.} and the energy level displacement of the ground state
of pionic deuterium \cite{AI5}.

The existence of the metastable resonance of the $np$ pair in the
spin--singlet state with a mass in the vicinity of the threshold mass
of the $np$ pair and the possibility to treat this metastable state as
a Cooper $np$--pair has been suggested by Ericson \cite{TE04} (see
also \cite{TE83}).

In this letter we follow a suggestion by Ericson \cite{TE04} and treat
the $np$ pair in the spin--singlet state as a Cooper pair in a way
analogous to the NNJL model \cite{AI1}--\cite{AI5}. The local
interpolating field operator of the $np$ pair in the spin--singlet
state we define as follows $S(x) = n^c(x)i\gamma^5 p(x)$ \cite{AI1},
where $n(x)$ and $p(x)$ are the interpolating field operators of the
neutron and the proton, respectively. The effective low--energy
Lagrangian, unstable with respect to the production of the metastable
resonant $np$ pair in the spin--singlet state and containing only the
fields of the neutron and the proton, we define as follows \cite{AI1}
\begin{eqnarray}\label{label2}
\hspace{-0.3in}{\cal L}^{\rm np}(x) &=&
\bar{n}(x)\,(i\gamma^{\mu}\partial_{\mu} - m_{\rm N})\,n(x)+
\bar{p}(x)\,(i\gamma^{\mu}\partial_{\mu} - m_{\rm N})\,p(x)\nonumber\\
\hspace{-0.3in}&+& \frac{g^2_{\rm S}}{4m^2_{\rm
N}}\,[\bar{p}(x)i\gamma^5 n^c(x)][n^c(x)i\gamma^5 p(x)] + \ldots,
\end{eqnarray}
where ellipses stand for low--energy interactions of the neutron and
the proton with other fields \cite{AI1}.  The coupling constant $g_S$
is related to the coupling constant $g_V$ of the NNJL
model\,\cite{AI1}\footnote{The coupling constant $g^2_V = 2\pi^2 m^2_N
Q_D = (11.3)^2$, where $Q_D = 0.286\,{\rm fm^2}$ is the electric
quadrupole moment of the deuteron, is responsible for the existence of
the deuteron as a Cooper $np$--pair \cite{AI1}.} as $g_S =
g_V/\sqrt{3}$ \cite{AI3}\,\footnote{Such a relation can be easily
obtained in the one--pion exchange approximation \cite{AI3}.}.

In order to introduce the interpolating local field $S(x)$ we should
linearalize the Lagrangian (\ref{label2}). Following the procedure
described in \cite{AI1} we get
\begin{eqnarray}\label{label3}
\hspace{-0.3in}{\cal L}^{\rm np}(x) &=&
\bar{n}(x)\,(i\gamma^{\mu}\partial_{\mu} - m_{\rm N})\,n(x) +
\bar{p}(x)\,(i\gamma^{\mu}\partial_{\mu} - m_{\rm N})\,p(x)
\nonumber\\ \hspace{-0.3in}&-& 4m^2_N\,S^{\dagger}(x)S(x) + g_{\rm
S}\,[n^c(x)i\gamma^5 p(x)]\,S^{\dagger}(x) + g_{\rm
S}\,[\bar{p}(x)i\gamma^5 n^c(x)]\,S(x) + \ldots.
\end{eqnarray}
For the derivation of the effective Lagrangian of the physical
$S$--field we should integrate over nucleon fields \cite{AI1}.

Following the procedure of the integration over the nucleon fields,
expounded in \cite{AI1}, we arrive at the effective Lagrangian of the
collective field $S(x) = n^c(x)i\gamma^5 p(x)$:
\begin{eqnarray}\label{label4}
\hspace{-0.3in}{\cal L}(x) = \Big(1 +
\frac{g^2_S}{8\pi^2}\,J_2(m_N)\Big)\partial_{\mu}S^{\dagger}(x)
\partial^{\mu}S(x) - \Big(4m^2_N -
\frac{g^2_S}{4\pi^2}\,J_1(m_N)\Big)\,S^{\dagger}(x)S(x),
\end{eqnarray}
where $J_1(m_N)$ and $J_2(m_N)$ are divergent integrals related by
\cite{AI1}
\begin{eqnarray}\label{label5}
J_1(m_N) = 2m^2_N J_2(m_N) = \frac{4}{3}\,\frac{\Lambda^3_D}{m_N} \sim
O(1/N_C).
\end{eqnarray}
The cut--off $\Lambda_D = 46\,{\rm MeV}$ \cite{AI1} restricts from
above 3--momenta of low--energy fluctuations of virtual neutron and
proton fields forming the physical metastable resonant $np$ state
\cite{AI1}. Making a renormalization of the wave function of the
collective field $S(x)$ \cite{AI1}, $S(x) \to Z^{1/2}S(x)$, where $Z =
1 + (g^2_S/8\pi^2)\,J_2(m_N)$ is the wave function renormalization
constant, we obtain the following effective Lagrangian:
\begin{eqnarray}\label{label6}
{\cal L}(x) = \partial_{\mu}S^{\dagger}(x) \partial^{\mu}S(x) -
m^2_S\,S^{\dagger}(x)S(x).
\end{eqnarray}
The mass of the field $S(x)$ is equal to
\begin{eqnarray}\label{label7}
m_S = Z^{-1/2}\sqrt{4m^2_N - \frac{g^2_S}{4\pi^2}\,J_1(m_N)} = 2m_N -
\frac{g^2_S}{4\pi^2}\,m_N J_2(m_N),
\end{eqnarray}
where we have used the relation (\ref{label5}).

Following Ref.\cite{AI1} we define the binding energy of the $np$ pair
in the spin--singlet state, described by the local interpolating field
$S(x)$. It reads
\begin{eqnarray}\label{label8}
\varepsilon_S = \frac{g^2_S}{4\pi^2}\,m_N J_2(m_N) =
\frac{g^2_V}{18\pi^2}\,\frac{\Lambda^3_D}{m^2_N} = (0.079\pm
0.012)\,{\rm MeV},
\end{eqnarray}
where $\pm 0.012$ is a theoretical accuracy of the approach
\cite{AI1}--\cite{AI5}.

The theoretical value of the binding energy $\varepsilon_S = (0.079\pm
0.012)\,{\rm MeV}$ agrees well with the binding energy
$\varepsilon_{{^1}{\rm S}_0} = 0.074\,{\rm MeV}$,
defined by the experimental S--wave scattering length of $np$
scattering in the spin--singlet state.

Using Eq.(\ref{label8}) we can compute the S--wave scattering length
of $np$ scattering in the spin--singlet state. Following Landau and
Lifshitz \cite{LL65}, we obtain
\begin{eqnarray}\label{label9}
a_S = - \frac{1}{\sqrt{\varepsilon_S m_N}} = (- 23.00 \pm 1.74)\,{\rm
fm}.
\end{eqnarray}
The theoretical value of $a_S$ agrees well with the experimental one
$a_{np} = (-\,23.75\pm 0.01)\,{\rm fm}$.

Thus, we have shown that, as it has been suggested by Ericson
\cite{TE04}, the $np$ pair in the spin--singlet state can be treated
as a Cooper $np$--pair with mass to be very close to the threshold
mass $2m_N$. The difference $\varepsilon_S$ between the threshold mass
and the mass of the Cooper $np$--pair mass can be identified with the
binding energy of the {\it virtual level} \cite{LL65}. The theoretical
value $\varepsilon_S = (0.079\pm 0.012)\,{\rm MeV}$ agrees well with
the the binding energy $\varepsilon_{{^1}{\rm S}_0} = 0.074\,{\rm
MeV}$ of the {\it virtual level} defined by the experimental value
$a_{np} = (-\,23.75\pm 0.01)\,{\rm fm}$ of the S--wave scattering
length of $np$ scattering in the spin--singlet state.

We are grateful to Torleif Ericson for the formulation of the problem
and numerous helpful discussions. Natalia Troitskaya and Andrei Ivanov
thank Torleif Ericson and Magda Ericson and TH--division of CERN for
kind hospitality extended to them during the period of their stay at
CERN, when this work was begun.

\end{document}